\documentclass[aps,pre,twocolumn,amsmath,amssymb,superscriptaddress,showpacs,longbibliography]{revtex4-1}

\usepackage{graphicx}
\usepackage{dcolumn}
\usepackage{bm}
\usepackage{mathrsfs}

\usepackage{natbib}
\usepackage[text={7in,9.5in},centering]{geometry}

\usepackage{hyperref}
\usepackage{url}

\usepackage{epsfig}
\usepackage{amsmath}
\usepackage{amssymb}
\usepackage{textcomp}

\usepackage{color}
\usepackage{float}

\usepackage{stackengine}
\usepackage{verbatim}

\begin{document}

\title{Approximating nonbacktracking centrality and localization phenomena in large networks}

\author{G. Tim\'ar}
 \email{gtimar@ua.pt}
 \affiliation{Departamento de F\'\i sica da Universidade de Aveiro \& I3N, Campus Universit\'ario de Santiago, 3810-193 Aveiro, Portugal}

\author{R. A. da Costa}
 \affiliation{Departamento de F\'\i sica da Universidade de Aveiro \& I3N, Campus Universit\'ario de Santiago, 3810-193 Aveiro, Portugal}

\author{S. N. Dorogovtsev}
 \affiliation{Departamento de F\'\i sica da Universidade de Aveiro \& I3N, Campus Universit\'ario de Santiago, 3810-193 Aveiro, Portugal}

\author{J. F. F. Mendes}
 \affiliation{Departamento de F\'\i sica da Universidade de Aveiro \& I3N, Campus Universit\'ario de Santiago, 3810-193 Aveiro, Portugal}

\date{\today}

\begin{abstract}
Message-passing theories have proved to be invaluable tools in studying percolation, non-recurrent epidemics and similar dynamical processes on real-world networks. At the heart of the message-passing method is the nonbacktracking matrix whose largest eigenvalue, the corresponding eigenvector, and the closely related nonbacktracking centrality play a central role in determining how the given dynamical model behaves. Here we propose a degree-class-based method to approximate these quantities using a smaller matrix related to the joint degree-degree distribution of neighbouring nodes. Our findings suggest that in most networks degree-degree correlations beyond nearest neighbour are actually not strong, and our first-order description already results in accurate estimates, particularly when message-passing itself is a good approximation to the original model in question, that is when the number of short cycles in the network is sufficiently low. We show that localization of the nonbacktracking centrality is also captured well by our scheme, particularly in large networks. Our method provides an alternative to working with the full nonbacktracking matrix in very large networks where this may not be possible due to memory limitations.
\end{abstract}

\maketitle

\section{Introduction}
\label{sec1}

A lot of recent scientific effort has been aimed at understanding how dynamical processes running on top of complex networks are affected by the underlying network structure. A large and relevant subset of dynamical processes can be accurately approximated by the message-passing method (also known as cavity method), where it is assumed that the contribution of a node $j$ to the behaviour of a neighbouring node $i$ is completely determined by the contribution of the neighbours of $j$, excluding $i$. This approximation is appropriate to study percolation \cite{karrer2014percolation, radicchi2015predicting} and spreading processes where a node may be activated at most once, as is the case in the SIR (susceptible, infected, recovered or removed) model of non-recurrent epidemics \cite{karrer2010message}. Message-passing methods disregard backtracking propagation, therefore their linearized version is described by the nonbacktracking (or Hashimoto) matrix \cite{hashimoto2014automorphic, martin2014localization} instead of the adjacency matrix. The nonbacktracking (NB) matrix $\mathbf{H}$ is a $2L \times 2L$ nonsymmetric matrix ($L$ being the number of links in the network) whose elements are indexed by directed links $i \leftarrow j$, instead of nodes. It is defined as $H_{i \leftarrow j, k \leftarrow l} = \delta_{j,k} (1 - \delta_{i,l})$, where $\delta$ is the Kronecker symbol. In Ref. \cite{timar2017nonbacktracking} it was shown that message-passing equations treat any finite loopy network as a well-defined infinite locally treelike network that preserves all local structures of the original, as seen by a nonbacktracking walker. This structure is encoded in the nonbacktracking matrix of the graph.

The key advantage of the NB matrix compared to the adjacency matrix is that it suffers to a much lesser degree from localization of the eigenvectors on large hubs \cite{martin2014localization}, due to the prohibition of backtracking. This circumstance has made it a useful tool in spectral community detection methods \cite{krzakala2013spectral, bordenave2015non}. The NB matrix has been used to design optimal percolation and node immunization strategies \cite{morone2015influence, morone2016collective, torres2020node}, identify influential spreaders \cite{radicchi2016leveraging, min2018identifying}, and estimate the time an epidemic takes to reach individual nodes in a network \cite{moore2020predicting}. The relevant quantity in most of these applications is the nonbacktracking centrality (NBC) of a node, defined as $x_i = \sum_{j \in \mathcal{N}_i} v_{i \leftarrow j}$, where $\mathcal{N}_i$ denotes the set of node $i$'s neighbours and $v_{i \leftarrow j}$ is the component of the principal eigenvector (PEV) of the NB matrix corresponding to the directed link $i \leftarrow j$. The largest eigenvalue (LEV) of the NB matrix plays the role of an effective ``branching number'', which determines the percolation or SIR epidemic threshold in the message-passing approximation of these processes \cite{karrer2014percolation, karrer2010message, timar2017nonbacktracking}. The NBC of a given node is proportional to the probability of belonging to the giant component (or of being infected in an SIR epidemic) close to the transition threshold. For this reason it is of great importance to know which nodes have the highest NBC, i.e., what group of nodes contribute most to the PEV of the NB matrix. Localization of the adjacency matrix PEV and its consequences for dynamical models such as the SIS (susceptible, infected, susceptible) epidemic model have been studied in detail, see, for example, Refs. \cite{goltsev2012localization, martin2014localization, pastor2018eigenvector}. In Refs. \cite{kawamoto2016localized, pastor2020localization} it has been suggested that the PEV of the NB matrix may also become localized, although not on individual hubs, but rather, on densely connected small subgraphs such as the highest $k$-core of the network or a group of ``overlapping hubs''. Using these findings, an estimate for the LEV of the NB matrix was given (see Ref. \cite{pastor2020localization}) which was found to be a strong improvement over the mean branching $\langle k^2 \rangle / \langle k \rangle - 1$.

We explore the possibility of approximating the LEV of the NB matrix and the NBC of nodes in a network considering only nearest-neighbour degree-degree correlations, i.e., substituting a given network with an infinite random network that has a joint degree-degree distribution $P(k,k')$ identical to the original network in question. Such an approximation is described by a matrix whose number of rows is equal to the number of different degrees in the network, which may be significantly smaller than the number of links. Therefore, if found to be accurate, such a degree-class-based approximation may be useful to estimate the percolation or epidemic threshold and individual NBC values of nodes in cases where the network in question is too large to be easily studied using the full NB matrix. We find that such a degree-based approximation indeed works well, and the relevant matrix is closely related to the branching matrix used in Refs. \cite{boguna2003epidemic, goltsev2008percolation}. Additionally we observe that localization of the NBC, quantified by the inverse participation ratio (IPR) is also reproduced fairly accurately in our method, lending more credence to the validity of a degree-based approximation.

\section{Two matrices describing correlated networks}
\label{sec2}

We discuss two related matrices that represent the same infinite maximally random network with given nearest neighbour degree-degree correlations described by the joint degree-degree distribution $P(k,k')$. We will use these matrices in Section \ref{sec3} to write approximations for the LEV of the NB matrix and the mean NBC of nodes of degree $k$.

\subsection{The branching matrix: percolation and SIR epidemics in correlated networks}
\label{sec21}

The branching matrix $\mathbf{B}$, defined as

\begin{align}
B_{k,k'} = (k'-1) P(k'|k),
\label{eq21.05}
\end{align}

\noindent
has been used to study SIR epidemics \cite{boguna2003epidemic} and percolation \cite{goltsev2008percolation} in random networks with only nearest neighbour degree-degree correlations. This matrix emerges by considering the probability $y_k$ that a random edge emanating from a node of degree $k$ leads to a finite component. Using the locally treelike property of an infinite random correlated network we can write the recursive equation

\begin{align}
y_k = \sum_{k'} P(k'|k) y_{k'}^{k'-1},
\label{eq21.10}
\end{align}

\noindent
where $P(k'|k)$ is the probability that a randomly chosen link has an end node of degree $k'$ given that the other end node is of degree $k$.
Assuming that $P(k'|k)$ is such that we are close to the percolation threshold we can write $a_k = 1 - y_k \ll 1$ and keep only terms linear in $a_k$,

\begin{align}
a_k = \sum_{k'} (k'-1) P(k'|k) a_{k'},
\label{eq21.20}
\end{align}

\noindent
or in vector form,

\begin{align}
\mathbf{a} = \mathbf{B} \mathbf{a},
\label{eq21.30}
\end{align}

\noindent
where the matrix $\mathbf{B}$ is the branching matrix defined in Eq. (\ref{eq21.05}).
Equation (\ref{eq21.30}), with the Perron-Frobenius theorem, implies that the LEV of matrix $\mathbf{B}$, at the percolation threshold, is $\lambda_1^{(\mathbf{B})} = 1$. Thus the quantity $\lambda_1^{(\mathbf{B})}$ is an effective branching in such correlated networks \cite{goltsev2008percolation}. From Eq. (\ref{eq21.30}) we also learn that close to the percolation threshold the probabiliy $a_k$ that a random link emanating from a node of degree $k$ leads to infinity is proportional to $v_k$, the appropriate component of the PEV of matrix $\mathbf{B}$. The probability that a node of degree $k$ belongs to the giant component is thus proportional to $k v_k$, and the sum of this probability over all nodes of degree $k$ is proportional to $kP(k)v_k$.

\subsection{The expansion matrix: nonbacktracking expansion of correlated networks}
\label{sec22}

\begin{figure}[H]
\centering
\includegraphics[width=6cm,angle=0.]{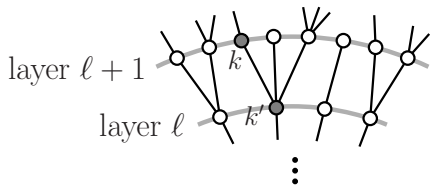}
\caption{Schematic representation of two successive layers in the expansion of a random network with nearest neighbour degree-degree correlations specified by a joint degree-degree distribution $P(k,k')$.}
\label{fig:schematic}
\end{figure}

A different way of obtaining a description of correlated networks is by following the ideas of Ref. \cite{timar2017nonbacktracking} according to which the nonbacktracking expansion is defined.
Let us build the expanding neighbourhood (a tree, layer by layer) of a random node in a network with only nearest neighbour degree-degree correlations. Below the percolation threshold, when $\lambda_1^{(\mathbf{B})} < 1$, such a construction will always end within a finite number of steps (since all nodes belong to finite components, i.e., they have finite neighbourhoods). When $\lambda_1^{(\mathbf{B})} > 1$, however, there is a nonzero probability $S$, that such a construction continues to infinity. (This happens when the randomly chosen starting node belongs to the giant component.) Using the quantities $y_k$ from Section \ref{sec21} we can write $S$ as

\begin{align}
S = 1 - \sum_k P(k) y_k^k,
\label{eq22.10}
\end{align}

\noindent
where $P(k) = ( \langle k \rangle / k ) \sum_{k'} P(k,k') $ is the degree distribution of the given correlated network.
Assuming that we can build an infinite expansion, let us calculate the relative frequency of nodes of degree $k$ on its ``boundary'' at infinity. In other words, we are interested in the relative frequency of nodes of degree $k$, i.e., the degree distribution, in an infinite local neighbourhood of a correlated network. Note that this is not the same as the distribution of degrees at the end of a random link, which is, in general, $kP(k)/\langle k \rangle$. (The two distributions are identical for uncorrelated networks, but not for correlated ones.) Let $n_k(\ell)$ denote the mean number of nodes of degree $k$ in layer $\ell$ of the expansion and let $n(\ell)$ be the mean total number of nodes in layer $\ell$ (see Fig. \ref{fig:schematic}). (The averages are taken over the randomness of the expansion process, governed by the joint degree-degree distribution $P(k,k')$.)

\noindent
Relating the quantities of layers $\ell$ and $\ell+1$ we can write

\begin{align}
n_k(\ell+1) = \sum_{k'}n_{k'}(\ell) (k'-1) P(k|k').
\label{eq22.20}
\end{align}

\noindent
Introducing $\eta(\ell) = n(\ell+1) / n(\ell)$ we get

\begin{align}
\eta(\ell) f_k(\ell+1) = \sum_{k'}f_{k'}(\ell) (k'-1) P(k|k'),
\label{eq22.30}
\end{align}

\noindent
where $f_k(\ell) = n_k(\ell) / n(\ell)$ is the relative mean number of nodes of degree $k$ on layer $\ell$. Assuming that the branching factor $\eta(\ell)$ and the relative mean numbers $f_k(\ell)$ converge to some constants for $\ell \to \infty$ we have the eigenvector equation

\begin{align}
\eta \mathbf{f} = \mathbf{E} \mathbf{f},
\label{eq22.40}
\end{align}

\noindent
with the \emph{expansion matrix} $\mathbf{E}$ defined as

\begin{align}
E_{k,k'} = (k'-1) P(k|k').
\label{eq22.50}
\end{align}

\noindent
The Perron-Frobenius theorem implies that $\eta$ and $\mathbf{f}$ are the LEV and PEV of the expansion matrix, respectively.

The branching and expansion matrices ($\mathbf{B}$ and $\mathbf{E}$) are two different ways of describing the same correlated network. The spectrum and eigenvectors of the two matrices are closely related, as is shown below in Section \ref{sec23}.

\subsection{Spectrum and eigenvectors of the two matrices}
\label{sec23}

Let $\lambda$ and $\mathbf{v}$ be an eigenvalue and corresponding eigenvector of matrix $\mathbf{B}$. Then

\begin{align}
\lambda v_k &= \sum_{k'} B_{k,k'} v_{k'}\\
&= \sum_{k'} (k'-1) P(k'|k) v_{k'}.
\label{eq23.20}
\end{align}

\noindent
Multiplying both sides by $k P(k)$ and using the relation $P(k'|k) k P(k) = P(k|k') k' P(k')$ \cite{boguna2002epidemic} we get

\begin{align}
\lambda k P(k) v_k &= \sum_{k'} (k'-1) P(k'|k) k P(k) v_{k'}\\
&= \sum_{k'} (k'-1) P(k|k') k' P(k') v_{k'},
\label{eq23.30}
\end{align}

\noindent
Introducing $\tilde{v}_k = k P(k) v_k$ we finally have

\begin{align}
\lambda \tilde{v}_k &= \sum_{k'} (k'-1) P(k|k') \tilde{v}_{k'}\\
&= \sum_{k'} E_{k,k'} \tilde{v}_{k'}.
\label{eq23.40}
\end{align}

\noindent
This means that the entire spectrum of the two matrices is identical, and if a vector with components $v_k$ is an eigenvector of $\mathbf{B}$ with eigenvalue $\lambda$, then the vector with components $k P(k) v_k$ is an eigenvector of $\mathbf{E}$ with the same eigenvalue $\lambda$.

\section{Approximating nonbacktracking centrality}
\label{sec3}

To approximate the NBC values of nodes in a given network, we construct the expansion matrix $\mathbf{E}$ (or the branching matrix $\mathbf{B}$) using the joint degree-degree distribution measured in the original network. To be precise, in a network consisting of $L$ links we count the number $L(k, k')$ of links connecting nodes of degrees $k$ and $k'$. The joint degree-degree distribution is then given as $P(k, k') = L(k, k') / L$ if $k = k'$ and $P(k, k') = L(k, k') / (2L)$ if $k \neq k'$. We may be more cautious and attempt to estimate the ``actual'' joint degree-degree distribution, assuming that the observed network is a single given realization of a certain stochastic generative process. This would however lead outside the scope of this paper, therefore, we simply count the number of links connecting nodes of given degrees. Consequently, in sparse networks the time complexity of our method is linear in system size.

We saw in Section \ref{sec21} that the components $v_k^{(\mathbf{B})}$ of the PEV of matrix $\mathbf{B}$ are proportional (close to the percolation threshold) to the probabilities that a link emanating from a node of degree $k$ leads to the giant component. The components $v_k^{(\mathbf{E})} = k P(k) v_k^{(\mathbf{B})}$ of the PEV of matrix $\mathbf{E}$, on the other hand, are proportional to the sum of probabilities of nodes of degree $k$ belonging to the giant component. In the message-passing scheme the NBC of a node is proportional to the probability of that node belonging to the giant component. Alternatively, in the nonbacktracking expansion of an arbitrary network, the NBC of a given node is equal to the relative frequency of replicas of that node on the boundary of the expansion at infinity (see Ref. \cite{timar2017nonbacktracking}). Similarly, in the expansion of a correlated network (Sec. \ref{sec22}) the relative frequency of nodes of degree $k$ on the boundary at infinity was found to be $v_k^{(\mathbf{E})}$.

The appropriate comparison is, therefore, between $v_k^{(\mathbf{E})}$ and the sum of NBC values of nodes of degree $k$,

\begin{align}
v_k^{(\mathbf{E})} \approx \sum_{i: k_i = k} x_i,
\label{eq3.10}
\end{align}

\noindent
where $x_i$ is the NBC of node $i$, $k_i$ denotes the degree of node $i$ and we assume the normalization $\sum_i x_i = \sum_k v_k^{(\mathbf{E})} = 1$. Equivalently we can write the approximation

\begin{align}
\langle x \rangle_k \approx \frac{v_k^{(\mathbf{E})}}{N P(k)} = \frac{k v_k^{(\mathbf{B})}}{N},
\label{eq3.20}
\end{align}

\noindent
where $\langle x \rangle_k$ denotes the mean NBC of nodes of degree $k$.
We make the assumption that a node's NBC is sufficiently well approximated by the mean NBC of its degree class, i.e.,

\begin{align}
x_i \approx \langle x \rangle_k \approx \frac{v_{k_i}^{(\mathbf{E})}}{N P(k_i)}.
\label{eq3.30}
\end{align}

\noindent
This (heterogeneous mean-field) approximation, arrived at by considering the nonbacktracking expansion of a correlated network, is derived more rigorously in the Appendix, using a methodology similar to that of Ref. \cite{fortunato2006approximating}. We show below, using a varied set of real-world example networks, that Eq. (\ref{eq3.30}) works fairly well for most and exceptionally well in certain cases, particularly when message-passing is itself a valid approximation of percolation-type phenomena.

The LEV of the NB matrix $\mathbf{H}$ plays the role of an effective branching in the given network and it determines the percolation (or SIR epidemic) threshold in the message-passing theory of these models. In our approximation we replace a given network with an infinite random network that has the same joint degree-degree distribution $P(k,k') = P(k|k') k' P(k') / \langle k \rangle$ as the original. Such an infinite random network is described by either matrix $\mathbf{E}$ or $\mathbf{B}$. An approximation to the LEV of the NB matrix is therefore simply given as

\begin{align}
\lambda_1^{(\mathbf{H})} \approx \lambda_1^{(\mathbf{E})} = \lambda_1^{(\mathbf{B})},
\label{eq3.40}
\end{align}

\noindent
see the Appendix for a derivation of this approximation. (If the original network is connected, then both matrices $\mathbf{E}$ and $\mathbf{B}$ are irreducible, which means that their largest eigenvalue is real and positive according to the Perron-Frobenius theorem.)
From here onwards we will refer to the approximations of Eqs. (\ref{eq3.20}) and (\ref{eq3.40}) as degree-based or expansion matrix approximations, although they could also be attributed to the branching matrix $\mathbf{B}$, as the two matrices contain the same information.

Recently, in Ref. \cite{pastor2020localization}, it was shown that the LEV of the NB matrix could be well approximated by the expression

\begin{align}
\mu = \textrm{max}(\mu^{\textrm{un}}, \mu^{\textrm{oh}}, \mu^{\textrm{core}}),
\label{eq3.50}
\end{align}

\noindent
where

\begin{align}
\mu^{\textrm{un}} = \frac{\sum_{ij} (k_i-1) A_{ij} (k_j-1)}{\sum_j k_j(k_j-1)}
\label{eq3.60}
\end{align}

\noindent
is an estimate based on the assumption that the network is uncorrelated. (The adjacency matrix is denoted by $\mathbf{A}$ and $k_i$ denotes the degree of node $i$.)
The quantities $\mu^{\textrm{oh}}$ and $\mu^{\textrm{core}}$ are the LEVs associated with the strongest ``overlapping hubs'' subgraph, and the highest $k$-core, respectively. The reason why these contributions must be dealt with separately is, as pointed out in Ref. \cite{pastor2020localization}, that such subgraphs are particularly sensitive to the given correlation patterns in a network, and their contribution is, in general, not included implicitly in $\mu^{\textrm{un}}$. Equation (\ref{eq3.50}) was found to be a significant improvement over the mean branching $\langle k^2 \rangle / \langle k \rangle - 1$, in approximating the LEV of the NB matrix.

\begin{figure}[H]
\centering
\includegraphics[width=\columnwidth,angle=0.]{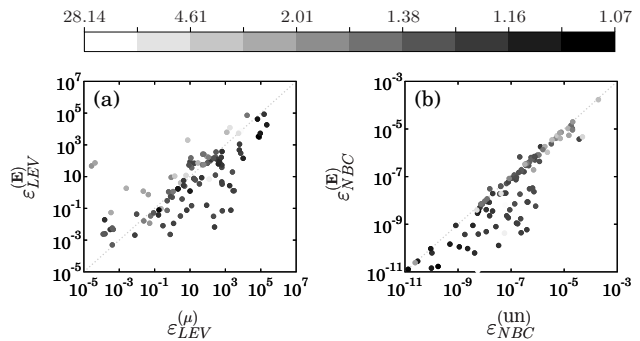}
\caption{(a) Error of the expansion matrix approximation to the LEV of the NB matrix as a function of the error of the approximation of Eq. (\ref{eq3.50}). (b) Error of the expansion matrix approximation to the NBC as a function of the error of the local approximation (Eq. (\ref{eq3.70})).
Each point on the panels shows the errors for one of 109 real-world networks. The errors are defined in Eqs. (\ref{eq3.65}), (\ref{eq3.66}) and (\ref{eq3.75}), (\ref{eq3.76}).
Points below the dashed grey lines in panels (a) and (b) have $\varepsilon_{LEV}^{(\mathbf{E})} < \varepsilon_{LEV}^{(\mu)}$ and $\varepsilon_{NBC}^{(\mathbf{E})} < \varepsilon_{NBC}^{(\textrm{un})}$, respectively. The color code in both panels corresponds to the ratio $\lambda_1^{(\mathbf{H})} / p_c^{-1}$, indicating the quality of the message-passing approximation to percolation.}
\label{fig:comparison}
\end{figure}

\noindent
In Fig. \ref{fig:comparison}(a) we compare our approximation [Eq. (\ref{eq3.40})] with that of Eq. (\ref{eq3.50}), for 109 real-world networks (see Table I in Supplemental Material) also considered in Ref. \cite{pastor2020localization}, featuring a variety of different sizes, clustering and correlation patterns. For the comparison we use, as a measure of the approximation error, the squared distances from the actual LEV of the NB matrix,

\begin{align}
\varepsilon_{LEV}^{(\mathbf{E})} &= \left( \lambda_1^{(\mathbf{H})} - \lambda_1^{(\mathbf{E})} \right)^2 \label{eq3.65} \\
\varepsilon_{LEV}^{(\mu)} &= \left( \lambda_1^{(\mathbf{H})} - \mu \right)^2 \label{eq3.66}
\end{align}

The expansion matrix LEV provides a better approximation in 77 of the 109 cases. More importantly, the expansion matrix approximation is consistently better when message-passing itself is a good approximation to percolation: the color code in Fig. \ref{fig:comparison} corresponds to the ratio of the LEV $\lambda_1^{(\mathbf{H})}$ of the NB matrix to the inverse percolation threshold $p_c^{-1}$ estimated via simulations (see Ref. \cite{pastor2020localization}). According to Ref. \cite{karrer2014percolation}, $\lambda_1^{(\mathbf{H})}$ is a good approximation to $p_c^{-1}$ in many empirical networks, and $\lambda_1^{(\mathbf{H})} \geq p_c^{-1}$ is strictly true in infinite networks. (The inequality was found to be true in all 109 empirical networks considered.)

\begin{figure}[H]
\centering
\includegraphics[width=\columnwidth,angle=0.]{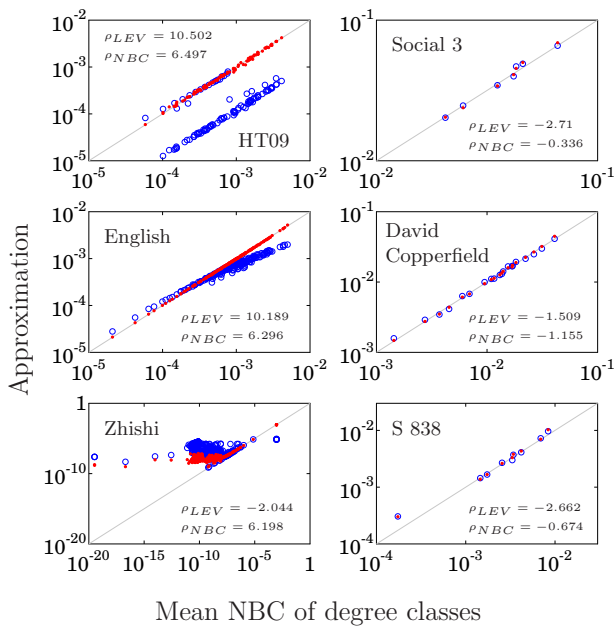}
\caption{Approximation of the mean NBC of nodes within degree classes using the expansion matrix approximation of Eq. (\ref{eq3.20}) (red dots) and the local estimate of Ref. \cite{pastor2020localization}, Eq. (\ref{eq3.70}) (open blue circles), averaged over the given degree class. (In each plot one marker corresponds to one degree class.) Left panels show results for the 3 highest, right panels for the 3 lowest values of the quantity $\rho_{\textrm{NBC}}$, Eq. (\ref{eq3.77}), indicating the 3 best and 3 worst cases (out of the 109) from the expansion matrix approximation viewpoint.}
\label{fig:best_worst}
\end{figure}

These results imply that whatever structure is responsible for the LEV of the NB matrix of the network, it is also captured implicitly in the correlated, degree-based, expansion matrix approximation. To what extent this holds true may be checked by looking at how well the mean NBC is approximated for individual degree classes. In Fig. \ref{fig:best_worst} we plot the expansion matrix approximation for the mean NBC of degree classes, as a function of the actual mean NBC (red dots), for 6 example networks.

\noindent

\noindent
The expansion matrix approximation of the mean NBC is compared with the uncorrelated approximation of Ref. \cite{pastor2020localization}, where a local estimate is given for the NBC of individual nodes,

\begin{align}
x_i^{\textrm{un}} \approx \frac{\sum_j A_{ij} (k_j-1)}{\sum_j k_j(k_j-1)}.
\label{eq3.70}
\end{align}

\noindent
The mean value of $x_i^{\textrm{un}}$ for degree classes is shown in Fig. \ref{fig:best_worst} as a function of the actual mean NBC (open blue circles). To quantify the quality of the two approximations we use the errors

\begin{align}
\varepsilon_{NBC}^{(\mathbf{E})} &= \sum_k^{k_{\textrm{max}}} \langle x \rangle_k \left( \langle x \rangle_k - \frac{v_{k_i}^{(\mathbf{E})}}{N P(k_i)} \right)^2, \label{eq3.75} \\
\varepsilon_{NBC}^{(\textrm{un})} &= \sum_k^{k_{\textrm{max}}} \langle x \rangle_k \Big( \langle x \rangle_k - \langle x \rangle_k^{\textrm{un}} \Big)^2, \label{eq3.76}
\end{align}

\noindent
which are the weighted sums of squared differences for the two approximations.
Degree classes of higher mean NBC generally play a bigger role in the underlying dynamics as described by the message-passing theory. It is therefore appropriate to use the mean NBC $\langle x \rangle_k$ as the weight in the above measure. To compare the two approximations we use the logarithm of the ratio of the respective errors,

\begin{align}
\rho_{\textrm{NBC}} = \ln  \left(  \varepsilon_{NBC}^{(\textrm{un})} /  \varepsilon_{NBC}^{(\mathbf{E})} \right).
\label{eq3.77}
\end{align}

\noindent
$\rho_{NBC} > 0$ indicates a smaller error for the expansion matrix approximation, while $\rho_{NBC} < 0$ indicates a smaller error for the approximation based on Eq. (\ref{eq3.70}). The 6 sample networks in Fig. (\ref{fig:best_worst}) were chosen to contain the 3 networks where the expansion matrix approximation worked best compared to the local one (highest $\rho_{NBC}$ values, left panels in Fig. (\ref{fig:best_worst})) and the 3 networks where it performed the worst (lowest $\rho_{NBC}$ values, right panels in Fig. (\ref{fig:best_worst})). We can observe only small differences between the two approximations in the worst cases, but striking differences in the best. Importantly, the NBCs of high degree nodes, which play a more important role, tend to be much better approximated by the expansion matrix. Equivalent figures for all 109 networks (showing similar trends) are presented in the Supplemental Material.

Fig. \ref{fig:comparison}(b) shows the error $\varepsilon_{\textrm{NBC}}^{(\mathbf{E})}$ as a function of $\varepsilon_{\textrm{NBC}}^{(\textrm{un})}$ for all 109 networks. The expansion matrix approximation is better in all but 6 cases. Importantly it is often markedly better when the message-passing approximation is itself valid. It is interesting to note that the approximation to the NBC works well in almost all cases, even when the LEV is badly approximated (see Fig. \ref{fig:comparison}(a)).

\begin{figure}[H]
\centering
\includegraphics[width=\columnwidth,angle=0.]{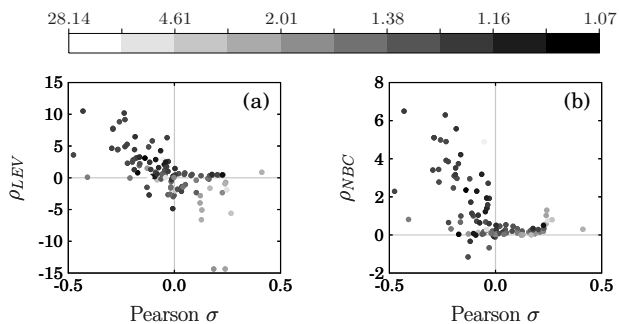}
\caption{Logarithmic approximation error ratios (a) $\rho_{\textrm{LEV}}$ and (b) $\rho_{\textrm{NBC}}$ as functions of the Pearson correlation coefficient $\sigma$ for nearest neighbour degree-degree correlations. The color code in both panels corresponds to the ratio $\lambda_1^{(\mathbf{H})} / p_c^{-1}$, indicating the quality of the message-passing approximation to percolation.}
\label{fig:rho_Pearson}
\end{figure}

It is worth analysing how the performance of the two approximations depends on particular degree-degree correlation patterns. The expansion matrix approximation accounts for nearest neighbour degree-degree correlations completely but assumes that there are no finite loops, which may be more problematic for certain types of networks than others.
Analogously to Eq. (\ref{eq3.77}) we define the following quantity to compare the two approximations to the LEV of the NB matrix,

\begin{align}
\rho_{\textrm{LEV}} = \ln  \left(  \varepsilon_{LEV}^{(\mu)} /  \varepsilon_{LEV}^{(\mathbf{E})} \right),
\label{eq3.78}
\end{align}

\noindent
which, again, is positive when the expansion matrix approximation is better. Fig. \ref{fig:rho_Pearson} shows the quantities $\rho_{\textrm{NBC}}$ and $\rho_{\textrm{LEV}}$ as functions of the Pearson correlation coefficient $\sigma$ (of nearest neighbour degrees) for the 109 networks considered. Considering the LEV approximation (Fig. \ref{fig:rho_Pearson}(a)) a clear trend is seen according to which the expansion matrix approximation works better for disassortative networks, while the approximation of Eq. (\ref{eq3.50}) favors assortative networks. The reason why the latter works better for assortative networks can be mostly attributed to the LEV of the highest $k$-core, which is explicitly included in Eq. (\ref{eq3.50}). Strongly assortative networks are expected to contain many short loops among the highest-degree nodes, most of which belong to the highest $k$-core, which in turn tends to dominate the LEV of the NB matrix. This feature is missed in the expansion matrix approximation, where a locally treelike structure is assumed. It is important to note, however, that for the very same reason message-passing theory itself is not a valid approximation in most of these assortative networks. Conversely, in the cases where message-passing is valid, the expansion matrix approximation is generally better, often markedly better. A similar trend can be seen in the case of the NBC approximation (Fig. \ref{fig:rho_Pearson}(b)), where the expansion matrix approximation tends to strongly dominate for disassortative networks. For the NBC approximation, interestingly, also for assortative networks the expansion matrix approximation appears to be better, or at least as good as the local approximation.

\noindent
The failure of the local approximation to correctly estimate the NBC in disassortative networks stems from the fact that the NBC of low degree nodes tends to be overestimated due to hubs in their immediate neighbourhood. The NBC of higher degree nodes is underestimated as a consequence. The expansion matrix method provides a reliable approximation for all degree classes in such networks. The obvious advantage of this method, compared to the local one, is that it is self-referential, i.e., it takes the entire network into account to determine the estimate for the mean NBC of degree classes, similarly to the message-passing algorithm, only it does so in a course-grained manner, circumventing the necessity to have access to the full NB matrix.

\begin{figure}[H]
\centering
\includegraphics[width=\columnwidth,angle=0.]{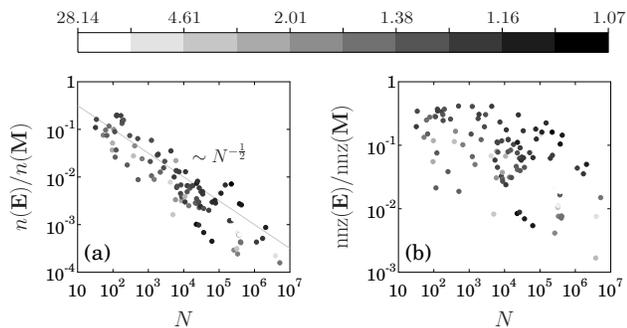}
\caption{Comparison of the size of the matrices $\mathbf{E}$ and $\mathbf{M}$ as a function of network size $N$. The ratio of the number of rows as a function of $N$ is presented in panel (a), the ratio of the number of nonzero elements as a function of $N$ is presented in panel (b). The color code in both panels corresponds to the ratio $\lambda_1^{(\mathbf{H})} / p_c^{-1}$, indicating the quality of the message-passing approximation to percolation.}
\label{fig:matsize}
\end{figure}

The quality of these findings indicates that whatever structural property of a network is responsible for determining the LEV of the NB matrix, nodes of identical degrees generally have similar roles, therefore they can be treated parsimoniously as a degree-class, if nearest-neighbour degree-degree correlations are taken into account. For large networks this may be a significant simplification and reduction in computer memory requirement. The NBCs in a given network can be calculated by first obtaining the PEV of the $2N \times 2N$ matrix

\begin{align}
\mathbf{M} =
\begin{pmatrix}
 \mathbf{A} & \mathbf{I} - \mathbf{D} \\
 \mathbf{I} & \mathbf{0}
\end{pmatrix}.
\label{eq3.80}
\end{align}

\noindent
where $\mathbf{A}$ is the adjacency matrix, $\mathbf{I}$ is the identity matrix and $\mathbf{D}$ is the ``degree matrix'' whose elements are $D_{ij} = \delta_{ij} k_i$. The NBC values $x_i$ correspond to the first $N$ components of the PEV of matrix $\mathbf{M}$ \cite{martin2014localization}. The number of rows in matrix $\mathbf{M}$ is $2N$, whereas the number of rows in the expansion matrix $\mathbf{E}$ is the number $n$ of different node degrees present in the network. The latter can be much smaller than the former for large networks. Figure \ref{fig:matsize}(a) shows that the ratio of the number of rows for the two matrices tends to decay with network size slightly faster than $N^{-1/2}$. This is a consequence of the fact that $n(\mathbf{E})$ is upper bounded by $k_{\textrm{max}}$ which is typically of the order of $N^{1/2}$ \cite{dorogovtsev2008critical}. For computational purposes however, what matters more than the number of rows is the number of nonzero elements in these matrices, $\textrm{nnz}(\mathbf{E})$ and $\textrm{nnz}(\mathbf{M})$, respectively. As can be seen in Fig. \ref{fig:matsize}(b) the ratio of this quantity for the two matrices does not decay as strongly as the ratio of the number of rows, but a decay is still evident.

\section{Localization of nonbacktracking centrality}
\label{sec4}

It is well established that the PEV of the adjacency matrix may become localized on hubs and their neighbouring nodes \cite{goltsev2012localization, castellano2012competing, martin2014localization, pastor2018eigenvector}. In particular, if the highest degree, $k_{\textrm{max}}$, in the network is larger than $( \langle k^2 \rangle / \langle k \rangle )^2$, then the PEV of the adjacency matrix is localized on this hub and the LEV is given by $\sqrt{k_{\textrm{max}}}$. (Otherwise the PEV is effectively localized on the highest $k$-core \cite{pastor2016distinct}.) This has significant consequences for recurrent epidemic models such as the SIS model, where it has been shown that in the quenched mean-field approximation the epidemic threshold coincides with the inverse of the LEV of the adjacency matrix \cite{wang2003epidemic}. Hubs in the SIS model, therefore have a special role in initiating disease spreading and maintaining an endemic state. For percolation and non-recurrent epidemics, for which the NB matrix and the NBC are the relevant quantities, hubs lose their special role, although not completely. Contrary to the case of the adjacency matrix PEV, independent hubs cannot be centers of localization of the NBC \cite{martin2014localization}. However, as shown recently in Refs. \cite{kawamoto2016localized, pastor2020localization}, the NBC may still become localized on high-degree nodes when they are supported by other high-degree nodes, either directly (in a densely connected subgraph, e.g., the highest $k$-core) or indirectly (in an ``overlapping hubs'' structure, where a group of high-degree nodes share the same neighbours).

Here we demonstrate that the expansion matrix approximation can also capture this localization phenomenon, which is consistent with the high quality of the NBC and LEV estimates. We quantify the localization of the NBC using the inverse participation ratio (IPR),

\begin{align}
Y_4 = \frac{ \sum_i x_i^4 }{ \left( \sum_i x_i^2 \right)^2 }.
\label{eq4.20}
\end{align}

\noindent
(The normalization $\sum_i x_i^2 = 1$ is often used.)
The quantity $Y_4$ may be approximated by replacing each $x_i$ with $\langle x \rangle_{k_i}$, the mean NBC value of nodes of degree $k_i$,

\begin{align}
\tilde{Y_4} = \frac{ \sum_k N P(k) \langle x \rangle_k^4 }{ \left( \sum_k N P(k) \langle x \rangle_k^2 \right)^2 }.
\label{eq4.30}
\end{align}

\noindent
In the expansion matrix approximation [Eq. (\ref{eq3.20})] we have $\langle x \rangle_k \approx v_k^{(\mathbf{E})} / (N P (k))$, where $v_k^{(\mathbf{E})}$ are the components of the PEV of the expansion matrix. Our approximation to the IPR of the NBC is then

\begin{align}
Y_4^{(\mathbf{E})} = \frac{    \sum_k    \left( v_k^{(\mathbf{E})} \right)^4 / ( N P(k) )^3     }     { \left[ \sum_k   \left( v_k^{(\mathbf{E})} \right)^2 /  ( N P(k) )   \right]^2 }.
\label{eq4.40}
\end{align}

\noindent
We will compare this estimate with the one obtained by using the uncorrelated, local approximation of the NBC, Eq. (\ref{eq3.70}),

\begin{align}
Y_4^{\textrm{un}} = \frac{ \sum_i (x_i^{\textrm{un}})^4 }{ \left( \sum_i (x_i^{\textrm{un}})^2 \right)^2 }.
\label{eq4.50}
\end{align}

\begin{figure}[H]
\centering
\includegraphics[width=\columnwidth,angle=0.]{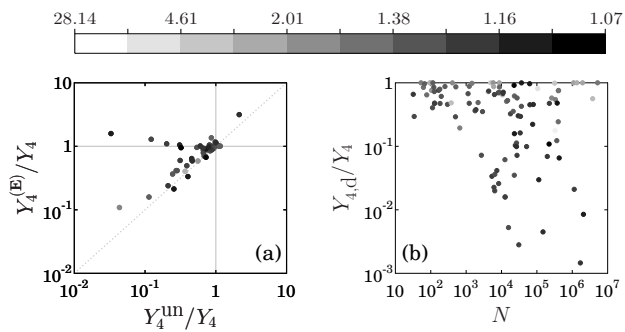}
\caption{(a) Comparison of the inverse participation ratio using the expansion matrix scheme and using the local approximation. Values relative to the true inverse participation ratio are shown on both axes. Dashed grey line corresponds to $Y_4^{(\mathbf{E})} = Y_4^{\textrm{un}}$. (b) Relative contribution of nodes of degenerate degrees to the inverse participation ratio $Y_4$, Eq. (\ref{eq4.20}), as a function of network size. The color code in both panels corresponds to the ratio $\lambda_1^{(\mathbf{H})} / p_c^{-1}$, indicating the quality of the message-passing approximation to percolation.}
\label{fig:IPR}
\end{figure}

\noindent
Note that Eq. (\ref{eq4.40}) is a degree-based, course-grained estimate, where each node is represented by the mean value of its degree class. Equation (\ref{eq4.50}), on the other hand, is a node-based estimate, i.e., the estimate of each node's individual NBC makes a contribution. It should be noted that Eq. (\ref{eq4.50}) is meant to be a good approximation primarily when the NBC is not localized on the highest $k$-core or overlapping hubs, i.e., when $\mu^{\textrm{un}}$ dominates in Eq. (\ref{eq3.50}). Data points comparing the approximations of Eqs. (\ref{eq4.40}) and (\ref{eq4.50}) for all such cases are shown in Fig. \ref{fig:IPR}(a). The expansion matrix approximation is better for 43 out of 55 networks. In the 12 cases where it is not, there is very little difference between the two approximations.

\noindent
These findings indicate that, for the purposes considered here, the role of a node is largely determined by its degree and the interaction between nodes of different degrees may be substituted by an averaged interaction between the respective degree classes.

Another important thing to consider is that although the NBC is not localized on independent hubs, the IPR is still dominated by high-degree nodes, due to their positions in densely connected subgraphs. Degree classes of high degrees have lower ``degeneracy'' (have fewer nodes in the degree class), therefore the degree-based approximation can be expected to work better. In particular, for nodes of the highest degrees, the degree classes are typically non-degenerate, and the expansion matrix describes their interconnections without any loss of information (compared to the NB matrix). Most of the information in the degree-based approximation is lost on low-degree classes that are highly degenerate, but these degree classes play a much smaller role in the localization phenomenon. We define $Y_{4,\textrm{d}}$ as the contribution of nodes of degenerate degree classes to the IPR,

\begin{align}
Y_{4,\textrm{d}} = \frac{ \sum\limits_{i:\textrm{degen.}} x_i^4 }{ \left( \sum_i x_i^2 \right)^2 },
\label{eq4.60}
\end{align}

\noindent
where the sum in the numerator is taken over nodes whose degrees are not unique in the network, that is, nodes that belong to degenerate degree classes.
In Fig. \ref{fig:IPR}(b) we plot the relative contribution of such degree classes to the IPR in all 109 networks. This contribution is often quite small, particularly for larger networks where the message-passing approximation is valid, meaning that the IPR is to a large extent dominated by nodes of unique degrees, whose interconnections are correctly described by the expansion matrix.

\section{Discussion and conclusions}
\label{sec7}

In this paper we propose an approximation to the NBC of nodes in a network and the related LEV of the NB matrix. Our approximation relies on the assumption that the given network behaves similarly to an infinite random network, without finite loops, that has the same nearest neighbour degree-degree correlations. Such correlated but otherwise uniformly random networks are described by a branching matrix---or equivalently, an expansion matrix---whose elements are related with the joint degree-degree distribution. The number of rows and columns in these matrices is given by the number of different degrees in the network, and hence they are generally much smaller than the NB matrix. The method we propose is degree-based, i.e., it is assumed that the NBCs of nodes are well approximated by the estimate of the mean NBC of their degree class. The estimates of the mean NBCs of degree classes are obtained by calculating the PEV of the branching or expansion matrix, and the estimate for the LEV of the NB matrix is given by the (identical) LEV of these two small matrices.

In spite of the fact that this method does not distinguish between nodes of identical degrees, the approximation for the mean NBC of degree classes and that for the LEV of the NB matrix are consistently better than existing local approximations to these quantities in networks where message-passing is a valid approximation to percolation-type phenomena. Importantly our method tends to approximate the NBC of high degree nodes better, which play a more important role in determining the LEV of the NB matrix. The small matrices in our method may be thought of as a ``compression'' of the NB matrix, where most of the information lost is on low-degree classes, which are generally strongly degenerate (contain many nodes). High-degree classes have much lower degeneracy, so the description of their interconnections remains true to the information contained in the original NB matrix. The connections between degree-classes that are non-degenerate (contain a single node) are described without any loss of information. In light of this fact it is understandable that the localization of the NBC, on subgraphs consisting of densely connected high-degree nodes, is also captured well in our approximation. That is, the localization of the PEV of the NB matrix is well traceable in the degree-based PEVs of the corresponding expansion or branching matrices. Our estimate of the inverse participation ratio is consistently better than that of the local approximation, despite the fact that in our method the NBCs of all nodes of identical degrees are considered equal.

The quality of our results also demonstrates that in most real-world networks considering only nearest neighbour degree-degree correlations is already sufficient for an accurate description---a description that is much better than the one resulting from assuming that the network is uncorrelated. An even more potent approximation may be achieved in principle by considering also triplet-wise (or even higher order) correlations, as opposed to only pair-wise. (A triplet is defined as a set of three nodes occupying the ends of two adjacent edges.) To construct the corresponding branching or expansion matrices, however, one would need to search through all the triplets (or higher-order structures) in the network. The number of triplets is dominated by the second moment of the degree distribution, therefore it may be very large for networks that possess fat-tailed degree distributions. Specifically, for scale-free networks the number of triplets (and of higher-order structures) is superlinear in system size if the degree distribution exponent is less than $3$. Thus computational complexity will, in most cases, constitute a barrier to considering triplet-wise (or higher order) correlations in large networks.

In social networks various characteristics of people may be correlated with degree, such as, e.g., age, profession and income status. Assuming that such correlations are known, our method provides a simple means of estimating the contribution of different groups of people in dynamical processes such as epidemics. These findings may help design preventative measures and vaccination strategies.

\section*{Acknowledgments}

We are grateful to Romualdo Pastor-Satorras and Claudio Castellano for providing us with the data for the 109 real-world networks.
This work was developed within the scope of the project i3N, UIDB/50025/2020 \& UIDP/50025/2020, financed by national funds through the FCT/MEC--Portuguese Foundation for Science and Technology. G.T. and R.A.d.C. were supported by FCT Grants No. CEECIND/03838/2017 and No. CEECIND/04697/2017.


\subsection*{Appendix: Derivation of expansion matrix approximation of nonbacktracking centrality}
\label{secA}
\setcounter{equation}{0}
\renewcommand{\theequation}{A\arabic{equation}}

The basic equations from which the NBC can be obtained are for directed links $i \leftarrow j$:

\begin{align}
v_{i \leftarrow j} = \lambda_1^{-1} \sum_{k \leftarrow l} H_{i \leftarrow j, k \leftarrow l} v_{k \leftarrow l},
\label{eqA.10}
\end{align}

\noindent
where $\mathbf{H}$ is the nonbacktracking matrix, and $\lambda_1$ and $\mathbf{v}$ are its largest eigenvalue and principal eigenvector, respectively. Let us define the quantity $\bar{v}(\vec{k})$ for directed links of degree class $\vec{k} = (k_1, k_2)$, as

\begin{align}
\bar{v}(\vec{k}) = \frac{1}{2L P(\vec{k})} \sum_{i \leftarrow j \in \vec{k}} v_{i \leftarrow j},
\label{eqA.20}
\end{align}

\noindent
where $2L$ is the number of directed links in the network and $P(\vec{k})$ is the probability that a uniformly randomly chosen directed link has end- and start-node degrees $k_1$ and $k_2$, respectively. The quantity $\bar{v}(\vec{k})$ is the mean $v_{i \leftarrow j}$ value over all links $i \leftarrow j$ where the degree of node $i$ is $k_1$ and the degree of node $j$ is $k_2$. Summing Eq. (\ref{eqA.10}) over all directed links of degree class $\vec{k}$ and dividing by $2L P(\vec{k})$ we have

\begin{align}
\bar{v}(\vec{k}) &= \frac{\lambda_1^{-1}}{2L P(\vec{k})} \sum_{i \leftarrow j \in \vec{k}} \sum_{k \leftarrow l} H_{i \leftarrow j, k \leftarrow l} v_{k \leftarrow l} \nonumber \\
&= \frac{\lambda_1^{-1}}{2L P(\vec{k})} \sum_{i \leftarrow j \in \vec{k}} \sum_{\vec{k}'} \sum_{k \leftarrow l \in \vec{k}'} H_{i \leftarrow j, k \leftarrow l} v_{k \leftarrow l}
\label{eqA.30}
\end{align}

\noindent
In the last equation we split the sum over directed links $k \leftarrow l$ into a sum over degree classes and a sum over links within degree classes. Now we make the mean-field assumption that all $v_{k \leftarrow l}$ values can be approximated by the corresponding mean value for the degree class,

\begin{align}
\bar{v}(\vec{k}) &= \frac{\lambda_1^{-1}}{2L P(\vec{k})} \sum_{i \leftarrow j \in \vec{k}} \sum_{\vec{k}'} \sum_{k \leftarrow l \in \vec{k}'} H_{i \leftarrow j, k \leftarrow l} \bar{v}(\vec{k}') \nonumber \\
&= \frac{\lambda_1^{-1}}{2L P(\vec{k})} \sum_{\vec{k}'} \bar{v}(\vec{k}') \left( \sum_{i \leftarrow j \in \vec{k}} \sum_{k \leftarrow l \in \vec{k}'} H_{i \leftarrow j, k \leftarrow l} \right).
\label{eqA.40}
\end{align}

\noindent
The quantity in large parentheses has a well-defined meaning: this is the number of $i_1 \leftarrow i_2 \leftarrow i_3$ directed triplets in the network where nodes $i_1$ and $i_2$ have degrees $k_1$ and $k_2$ and also, the nodes $i_2$ and $i_3$ have degrees $k_1'$ and $k_2'$. Or, more succinctly, this is the number of directed link junctions of type $\vec{k} \leftarrow \vec{k}'$. Let us approximate this by the number of such directed link junctions in a network with nearest neighbour degree-degree correlations described by a joint degree-degree distribution. This number is

\begin{align}
J_{\vec{k} \leftarrow \vec{k}'} = 2L P(\vec{k}') (k_2 - 1) P(k_1 | k_2) \delta_{k_2, k_1'},
\label{eqA.50}
\end{align}

\noindent
where $\delta_{k,k'}$ is the Kronecker delta. Plugging Eq. (\ref{eqA.50}) into Eq. (\ref{eqA.40}) and rearranging we have

{
\medmuskip=0mu
\thinmuskip=0mu
\thickmuskip=0mu
\begin{align}
2L P(\vec{k}) \bar{v}(\vec{k}) = \lambda_1^{-1} \sum_{\vec{k}'} 2L P(\vec{k}') \bar{v}(\vec{k}')  (k_2 - 1) P(k_1 | k_2) \delta_{k_2, k_1'}.
\label{eqA.60}
\end{align}
}

\noindent
Recall that $\bar{v}(\vec{k})$ is the mean $v_{i \leftarrow j}$ value of directed links of degree class $\vec{k}$. Then the left-hand side of Eq. (\ref{eqA.60}) is just the sum of such values. Denoting this sum by $g(\vec{k}) = g(k_1 \leftarrow k_2)$ we can write

{
\medmuskip=0mu
\thinmuskip=0mu
\thickmuskip=0mu
\begin{align}
g(k_1 \leftarrow k_2) = \lambda_1^{-1} \sum_{k_1' \leftarrow k_2'} g(k_1' \leftarrow k_2')  (k_2 - 1) P(k_1 | k_2) \delta_{k_2, k_1'},
\label{eqA.70}
\end{align}
}

\noindent
where we have switched from vectorial to component-wise notation for degree classes of directed links. Rearranging, we get

{
\medmuskip=0mu
\thinmuskip=0mu
\thickmuskip=0mu
\begin{align}
g(k_1 \leftarrow k_2) &= \lambda_1^{-1} (k_2 - 1) P(k_1 | k_2)  \sum_{k_1' \leftarrow k_2'} g(k_1' \leftarrow k_2') \delta_{k_2, k_1'} \nonumber \\
&= \lambda_1^{-1} (k_2 - 1) P(k_1 | k_2)  \sum_{k_2'} g(k_2 \leftarrow k_2').
\label{eqA.80}
\end{align}
}

\noindent
Recall that $g(k \leftarrow k')$ is the sum of $v_{i \leftarrow j}$ values of directed links within the degree class $k \leftarrow k'$. The sum of $v_{i \leftarrow j}$ values over all directed links incoming to nodes of degree class $k$ is then written as

\begin{align}
h_k = \sum_{k'} g(k \leftarrow k').
\label{eqA.90}
\end{align}

\noindent
Using Eq. (\ref{eqA.90}) and summing both sides of Eq. (\ref{eqA.80}) over $k_2$ we have

\begin{align}
h_{k_1} = \lambda_1^{-1} \sum_{k_2} (k_2 - 1) P(k_1 | k_2) h_{k_2}.
\label{eqA.100}
\end{align}

\noindent
Introducing the matrix $E_{k_1,k_2} = (k_2 - 1) P(k_1 | k_2)$ we finally have

\begin{align}
h_{k_1} = \lambda_1^{-1} \sum_{k_2} E_{k_1, k_2} h_{k_2},
\label{eqA.110}
\end{align}

\noindent
or in vector form

\begin{align}
\lambda_1 \mathbf{h} = \mathbf{E} \mathbf{h}.
\label{eqA.120}
\end{align}

\noindent
Equation (\ref{eqA.120}) is an eigenvector equation for the principal eigenvector of $\mathbf{E}$ (by the Perron-Frobenius theorem). The components of the vector $\mathbf{h}$ approximate the sum of NBCs of nodes of the corresponding degree class, and the LEV of $\mathbf{E}$ is an approximation to $\lambda_1$, the LEV of the NB matrix.


%

\end{document}



  

{\centering{\huge Supplemental Material to \emph{Approximating nonbacktracking centrality and localization phenomena in large networks}\par}\vspace{3ex}
	{\Large G. Tim\'ar, R. A. da Costa, S. N. Dorogovtsev and J. F. F. Mendes\par}\vspace{5ex}}

This Supplemental Material contains the set of plots comparing approximations of the nonbacktracking centrality in 109 empirical networks and the table with key characteristics of these networks and numbers quantifying the accuracy of different approximations to the nonbacktracking centrality.

\vskip 1cm

\noindent
\centering
{\Large Supplementary figures}

\begin{figure*}[h]
\centering
\caption{Approximation of the mean NBC of nodes within degree classes using the expansion matrix approximation of Eq. (18) (red dots) and the local estimate of Ref. [18], Eq. (24) (open blue circles), averaged over the given degree class. (In each plot one marker corresponds to one degree class.) The networks are identified by their index in Table \ref{table1}.}
\includegraphics[width=\textwidth,angle=0.]{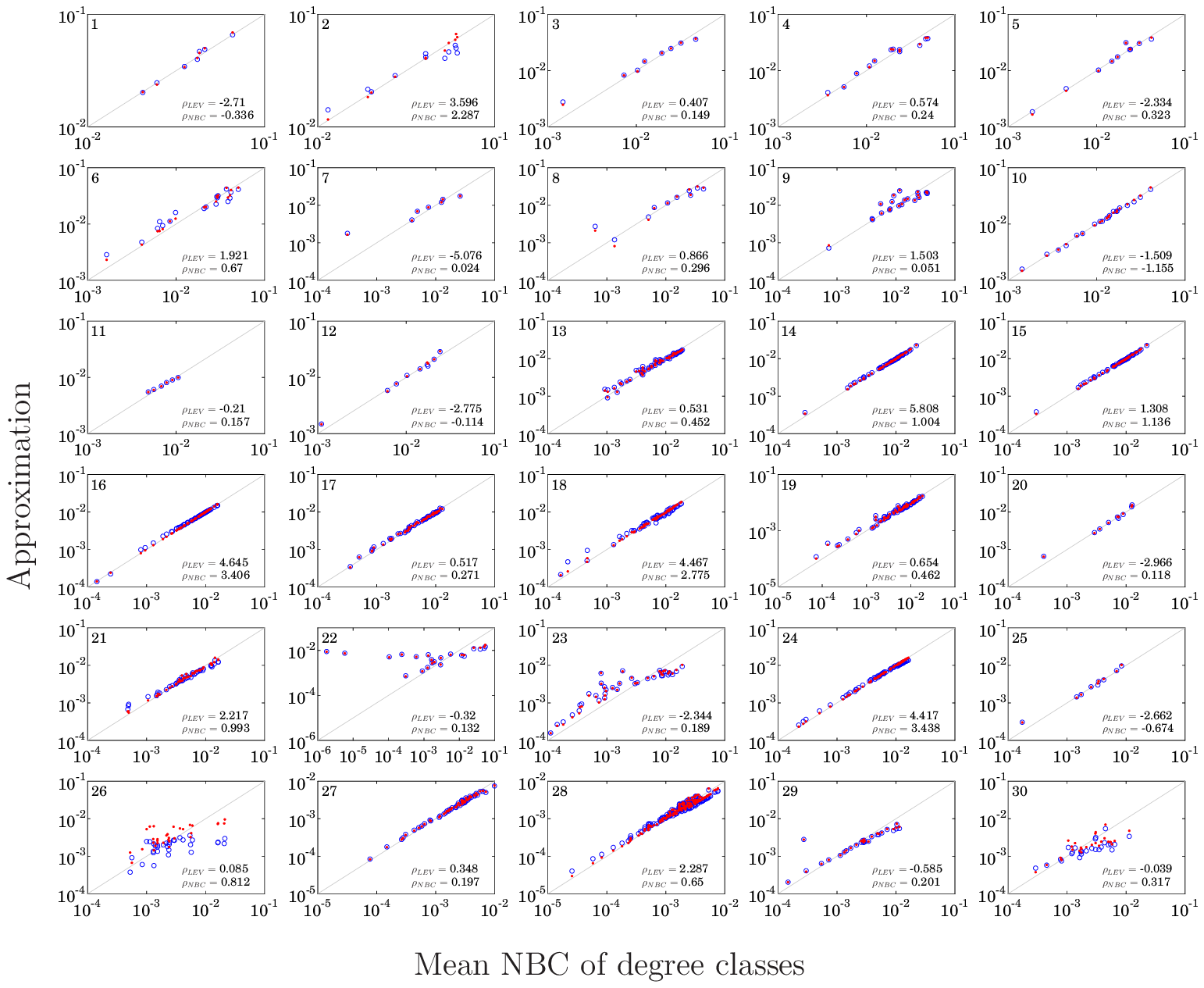}
\end{figure*}

\begin{figure*}[h]
\ContinuedFloat
\centering
\caption{}
\includegraphics[width=\textwidth,angle=0.]{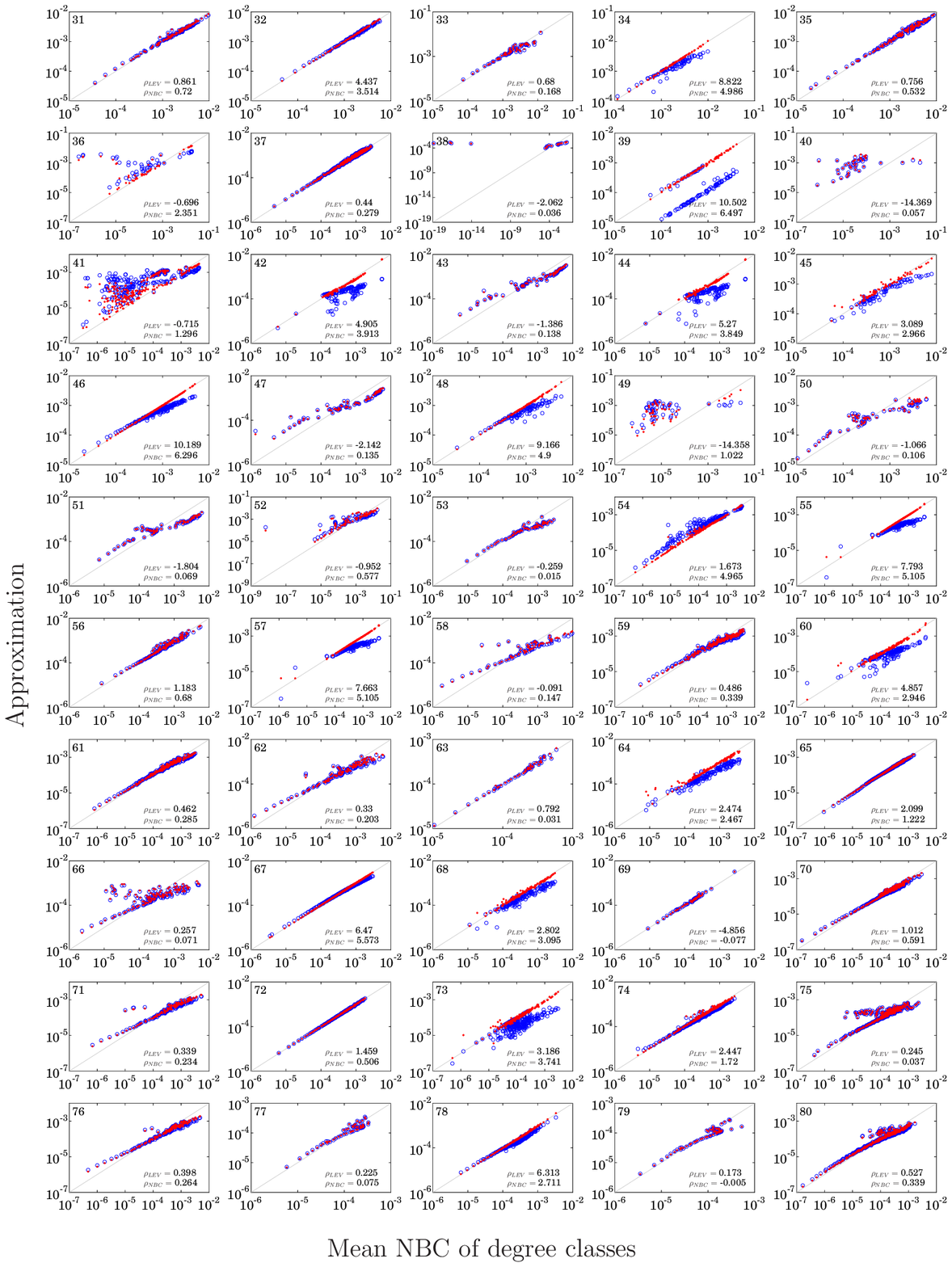}
\end{figure*}

\begin{figure*}[h]
\ContinuedFloat
\centering
\caption{}
\includegraphics[width=\textwidth,angle=0.]{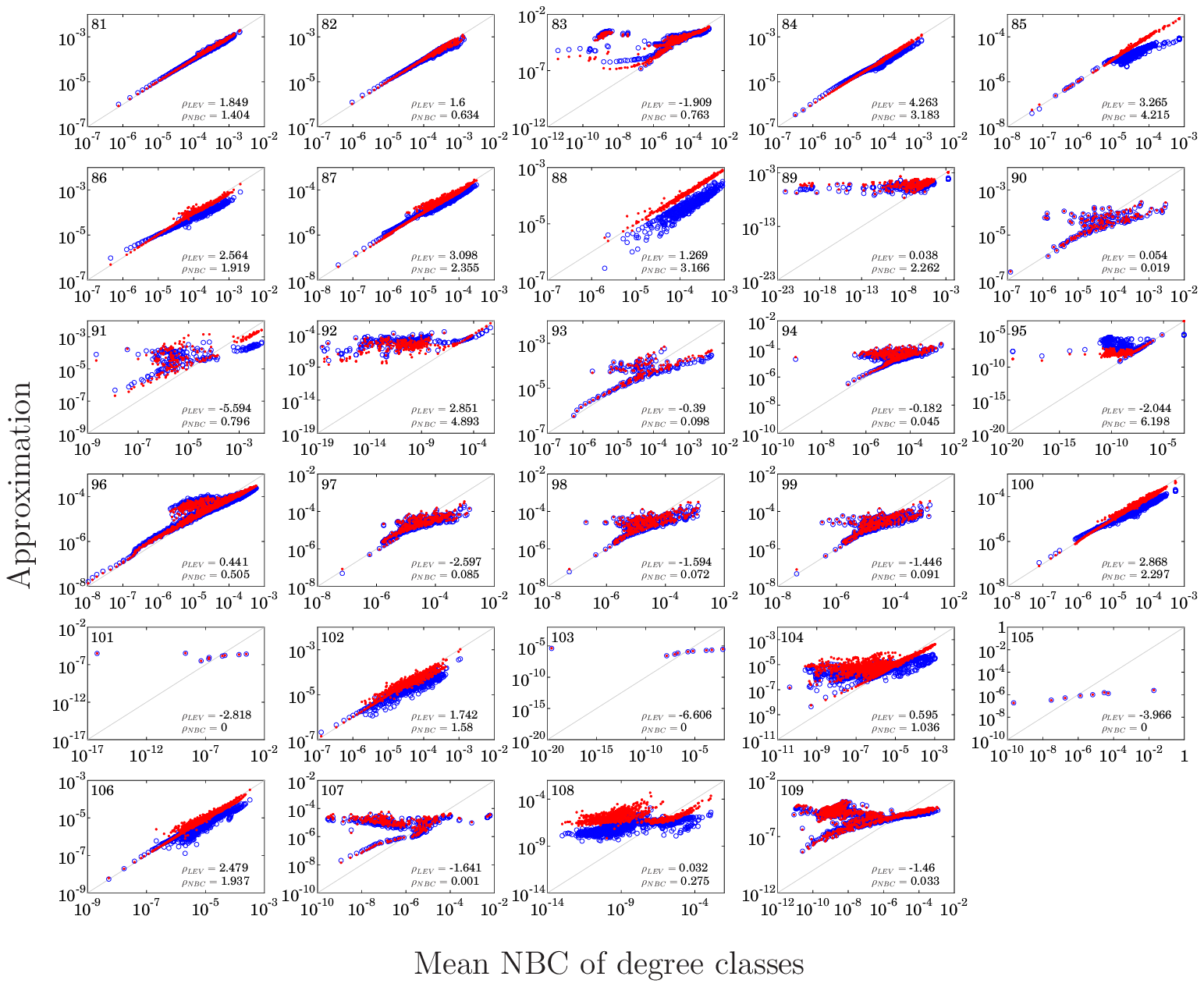}
\label{fig:ALL}
\end{figure*}

\clearpage

\noindent
{\Large Supplementary table}

\setlength{\tabcolsep}{4pt}
\setlength{\LTcapwidth}{\textwidth}
\begin{longtable*}{lllllllllllll}

\caption{Characteristics of the 109 real-world networks considered in this study. The values presented for each network are the number of nodes $N$; mean degree $\langle k \rangle$; maximum degree $k_{\textrm{max}}$; rank of matrix $\mathbf{E}$ (number of different degrees) $\textrm{rank}(\mathbf{E})$; LEV of the NB matrix $\mathbf{H}$, $\lambda_1^{(\mathbf{H})}$; LEV of matrix $\mathbf{E}$, $\lambda_1^{(\mathbf{E})}$; approximation of Eq. (20) to the LEV of the NB matrix, $\mu$; mean branching $\langle k^2 \rangle / \langle k \rangle-1$; the Pearson correlation coefficient $\sigma$; and the logarithmic ratios $\rho_{NBC}$ and $\rho_{LEV}$ describing the relative accuracy of the two approximations to the NBC of degree classes and to the LEV of the NB matrix.} \\

\label{table1} \\

     & Network   & $N$ & $\langle k \rangle$ & $k_{\textrm{max}}$ & $\textrm{rank}(\mathbf{E})$ & $\lambda_1^{(\mathbf{H})}$ & $\lambda_1^{(\mathbf{E})}$ & $\mu$ & $\frac{\langle k^2 \rangle}{\langle k \rangle}-1$ & $\sigma$ & $\rho_{NBC}$ & $\rho_{LEV}$ \\ \hline

\endfirsthead

     & Network   & $N$ & $\langle k \rangle$ & $k_{\textrm{max}}$ & $\textrm{rank}(\mathbf{E})$ & $\lambda_1^{(\mathbf{H})}$ & $\lambda_1^{(\mathbf{E})}$ & $\mu$ & $\frac{\langle k^2 \rangle}{\langle k \rangle}-1$ & $\sigma$ & $\rho_{NBC}$ & $\rho_{LEV}$ \\ \hline

\endhead
1.   & Social 3              & 32      & 5.00    & 13          & 7       & 4.74    & 4.81    & 4.76        & 4.94            & -0.119  & -0.336        & -2.710        \\
2.   & Karate club           & 34      & 4.59    & 17          & 11      & 5.29    & 5.38    & 4.75        & 6.77            & -0.476  & 2.287         & 3.596         \\
3.   & Protein 2             & 53      & 4.64    & 8           & 8       & 4.68    & 4.55    & 4.52        & 4.39            & 0.209   & 0.149         & 0.407         \\
4.   & Dolphins              & 62      & 5.13    & 12          & 12      & 5.99    & 5.81    & 5.75        & 5.81            & -0.044  & 0.240         & 0.574         \\
5.   & Social 1              & 67      & 4.24    & 11          & 10      & 4.36    & 4.43    & 4.38        & 4.25            & 0.103   & 0.323         & -2.334        \\
6.   & Les Miserables        & 77      & 6.60    & 36          & 18      & 10.75   & 10.48   & 10.04       & 11.06           & -0.165  & 0.670         & 1.921         \\
7.   & Protein 1             & 95      & 4.48    & 7           & 7       & 4.25    & 4.02    & 4.23        & 3.95            & 0.129   & 0.024         & -5.076        \\
8.   & E Coli transcription  & 97      & 4.37    & 10          & 10      & 5.34    & 5.03    & 4.86        & 4.41            & 0.412   & 0.296         & 0.866         \\
9.   & Political books       & 105     & 8.40    & 25          & 21      & 10.63   & 10.52   & 10.40       & 10.93           & -0.128  & 0.051         & 1.503         \\
10.  & David Copperfield     & 112     & 7.59    & 49          & 20      & 11.54   & 11.75   & 11.44       & 12.77           & -0.129  & -1.155        & -1.509        \\
11.  & College football      & 115     & 10.66   & 12          & 6       & 9.77    & 9.75    & 9.75        & 9.73            & 0.162   & 0.157         & -0.210        \\
12.  & S 208                 & 122     & 3.10    & 10          & 9       & 2.75    & 2.80    & 2.76        & 2.77            & -0.002  & -0.114        & -2.775        \\
13.  & High school 2011      & 126     & 27.13   & 55          & 49      & 32.85   & 32.30   & 32.13       & 31.79           & 0.083   & 0.452         & 0.531         \\
14.  & Bay Dry               & 128     & 32.91   & 110         & 51      & 38.91   & 38.86   & 38.04       & 39.50           & -0.104  & 1.004         & 5.808         \\
15.  & Bay Wet               & 128     & 32.42   & 110         & 50      & 38.44   & 38.40   & 38.53       & 39.11           & -0.112  & 1.136         & 1.308         \\
16.  & Radoslaw Email        & 167     & 38.92   & 139         & 65      & 59.43   & 59.54   & 58.35       & 63.46           & -0.295  & 3.406         & 4.645         \\
17.  & High school 2012      & 180     & 24.67   & 56          & 48      & 29.01   & 28.79   & 28.72       & 28.55           & 0.046   & 0.271         & 0.517         \\
18.  & Little Rock Lake      & 183     & 26.60   & 105         & 55      & 40.06   & 39.88   & 38.37       & 41.89           & -0.266  & 2.775         & 4.467         \\
19.  & Jazz                  & 198     & 27.70   & 100         & 62      & 38.82   & 38.11   & 37.83       & 37.64           & 0.020   & 0.462         & 0.654         \\
20.  & S 420                 & 252     & 3.17    & 14          & 9       & 2.89    & 2.94    & 2.90        & 2.91            & -0.006  & 0.118         & -2.966        \\
21.  & C Elegans neural      & 297     & 14.46   & 134         & 44      & 22.76   & 22.13   & 20.87       & 25.05           & -0.163  & 0.993         & 2.217         \\
22.  & Network Science       & 379     & 4.82    & 34          & 21      & 8.71    & 6.70    & 7.00        & 7.02            & -0.082  & 0.132         & -0.320        \\
23.  & Dublin                & 410     & 13.49   & 50          & 37      & 22.24   & 19.24   & 21.31       & 17.72           & 0.226   & 0.189         & -2.344        \\
24.  & US Air Transportation & 500     & 11.92   & 145         & 64      & 46.54   & 46.15   & 43.04       & 52.78           & -0.268  & 3.438         & 4.417         \\
25.  & S 838                 & 512     & 3.20    & 22          & 9       & 2.94    & 3.02    & 2.96        & 3.03            & -0.030  & -0.674        & -2.662        \\
26.  & Yeast transcription   & 662     & 3.21    & 71          & 32      & 6.50    & 5.61    & 5.57        & 12.51           & -0.410  & 0.812         & 0.085         \\
27.  & URV email             & 1133    & 9.62    & 71          & 48      & 19.27   & 18.51   & 18.37       & 17.69           & 0.078   & 0.197         & 0.348         \\
28.  & Political blogs       & 1222    & 27.36   & 351         & 144     & 72.56   & 70.70   & 66.72       & 80.26           & -0.221  & 0.650         & 2.287         \\
29.  & Air traffic           & 1226    & 3.93    & 34          & 26      & 7.48    & 6.43    & 6.70        & 6.36            & -0.015  & 0.201         & -0.585        \\
30.  & Yeast protein\_1      & 1458    & 2.67    & 56          & 26      & 5.05    & 3.98    & 4.00        & 6.13            & -0.210  & 0.317         & -0.039        \\
31.  & Petster hamster       & 1788    & 13.96   & 272         & 103     & 44.31   & 41.63   & 40.19       & 44.55           & -0.089  & 0.720         & 0.861         \\
32.  & UC Irvine             & 1893    & 14.62   & 255         & 114     & 46.25   & 45.98   & 43.70       & 54.64           & -0.188  & 3.514         & 4.437         \\
33.  & Yeast protein\_2      & 2172    & 6.05    & 215         & 55      & 18.54   & 16.95   & 16.31       & 18.79           & -0.055  & 0.168         & 0.680         \\
34.  & Japanese              & 2698    & 5.93    & 725         & 84      & 38.16   & 38.34   & 23.56       & 107.61          & -0.259  & 4.986         & 8.822         \\
35.  & Open flights          & 2905    & 10.77   & 242         & 124     & 61.33   & 58.55   & 57.28       & 54.84           & 0.049   & 0.532         & 0.756         \\
36.  & GR-QC 1993-2003       & 4158    & 6.46    & 81          & 65      & 44.44   & 40.99   & 42.00       & 16.98           & 0.639   & 2.351         & -0.696        \\
37.  & Tennis                & 4338    & 37.74   & 451         & 306     & 160.17  & 158.50  & 158.09      & 157.91          & 0.003   & 0.279         & 0.440         \\
38.  & US Power grid         & 4941    & 2.67    & 19          & 16      & 6.23    & 2.96    & 5.06        & 2.87            & 0.003   & 0.036         & -2.062        \\
39.  & HT09                  & 5352    & 6.91    & 1287        & 107     & 41.01   & 40.93   & 25.42       & 198.98          & -0.431  & 6.497         & 10.502        \\
40.  & Hep-Th 1995-1999      & 5835    & 4.74    & 50          & 39      & 17.01   & 10.13   & 17.00       & 8.12            & 0.185   & 0.057         & -14.369       \\
41.  & Reactome              & 5973    & 48.81   & 855         & 264     & 206.88  & 193.34  & 197.41      & 142.31          & 0.241   & 1.296         & -0.715        \\
42.  & Jung                  & 6120    & 16.43   & 5655        & 167     & 128.35  & 130.50  & 103.36      & 990.77          & -0.233  & 3.913         & 4.905         \\
43.  & Gnutella Aug 8 2002   & 6299    & 6.60    & 97          & 76      & 26.51   & 18.19   & 22.35       & 16.66           & 0.036   & 0.138         & -1.386        \\
44.  & JDK                   & 6434    & 16.68   & 5923        & 172     & 129.28  & 131.11  & 103.74      & 981.71          & -0.223  & 3.849         & 5.270         \\
45.  & AS Oregon             & 6474    & 3.88    & 1458        & 83      & 35.04   & 31.52   & 18.52       & 163.81          & -0.182  & 2.966         & 3.089         \\
46.  & English               & 7377    & 11.98   & 2568        & 190     & 104.34  & 104.62  & 59.17       & 319.70          & -0.237  & 6.296         & 10.189        \\
47.  & Gnutella Aug 9 2002   & 8104    & 6.42    & 102         & 73      & 26.56   & 17.30   & 23.39       & 15.82           & 0.033   & 0.135         & -2.142        \\
48.  & French                & 8308    & 5.74    & 1891        & 113     & 52.46   & 52.73   & 26.58       & 217.01          & -0.233  & 4.900         & 9.166         \\
49.  & Hep-Th 1993-2003      & 8638    & 5.74    & 65          & 55      & 30.01   & 21.46   & 30.00       & 11.99           & 0.239   & 1.022         & -14.358       \\
50.  & Gnutella Aug 6 2002   & 8717    & 7.23    & 115         & 67      & 20.47   & 14.46   & 16.94       & 13.40           & 0.052   & 0.106         & -1.066        \\
51.  & Gnutella Aug 5 2002   & 8842    & 7.20    & 88          & 68      & 21.58   & 14.28   & 18.62       & 13.79           & 0.015   & 0.069         & -1.804        \\
52.  & PGP                   & 10680   & 4.55    & 205         & 83      & 41.03   & 32.50   & 35.73       & 17.88           & 0.238   & 0.577         & -0.952        \\
53.  & Gnutella Aug 4 2002   & 10876   & 7.35    & 103         & 65      & 15.28   & 12.90   & 13.19       & 12.97           & -0.013  & 0.015         & -0.259        \\
54.  & Hep-Ph 1993-2003      & 11204   & 21.00   & 491         & 289     & 243.75  & 240.82  & 237.00      & 129.88          & 0.629   & 4.965         & 1.673         \\
55.  & Spanish               & 12643   & 8.70    & 5169        & 166     & 100.13  & 101.27  & 44.11       & 806.66          & -0.290  & 5.105         & 7.793         \\
56.  & DBLP citations        & 12495   & 7.93    & 709         & 122     & 38.06   & 35.58   & 33.58       & 42.77           & -0.046  & 0.680         & 1.183         \\
57.  & Spanish 2             & 12643   & 8.70    & 5169        & 166     & 100.13  & 101.27  & 47.63       & 806.66          & -0.290  & 5.105         & 7.663         \\
58.  & Cond-Mat 1995-1999    & 13861   & 6.44    & 107         & 69      & 23.14   & 15.67   & 16.00       & 12.54           & 0.157   & 0.147         & -0.091        \\
59.  & Astrophysics          & 14845   & 16.12   & 360         & 173     & 72.21   & 59.19   & 55.60       & 44.46           & 0.228   & 0.339         & 0.486         \\
60.  & Google                & 15763   & 18.85   & 11401       & 201     & 156.61  & 161.02  & 106.57      & 900.63          & -0.122  & 2.946         & 4.857         \\
61.  & AstroPhys 1993-2003   & 17903   & 22.00   & 504         & 234     & 92.54   & 80.79   & 77.74       & 64.70           & 0.201   & 0.285         & 0.462         \\
62.  & Cond-Mat 1993-2003    & 21363   & 8.55    & 279         & 122     & 35.80   & 27.51   & 26.02       & 21.47           & 0.125   & 0.203         & 0.330         \\
63.  & Gnutella Aug 25 2002  & 22663   & 4.83    & 66          & 46      & 9.38    & 9.10    & 8.96        & 9.75            & -0.173  & 0.031         & 0.792         \\
64.  & Internet              & 22963   & 4.22    & 2390        & 161     & 64.68   & 57.51   & 39.97       & 260.46          & -0.198  & 2.467         & 2.474         \\
65.  & Thesaurus             & 23132   & 25.69   & 1062        & 342     & 97.70   & 96.59   & 94.53       & 102.29          & -0.048  & 1.222         & 2.099         \\
66.  & Cora                  & 23166   & 7.70    & 377         & 129     & 29.28   & 20.61   & 19.42       & 22.68           & -0.055  & 0.071         & 0.257         \\
67.  & Linux mailing list    & 24567   & 12.88   & 2989        & 401     & 220.15  & 218.50  & 178.45      & 339.98          & -0.185  & 5.573         & 6.470         \\
68.  & AS Caida              & 26475   & 4.03    & 2628        & 158     & 59.41   & 53.25   & 34.41       & 279.24          & -0.195  & 3.095         & 2.802         \\
69.  & Gnutella Aug 24 2002  & 26498   & 4.93    & 355         & 52      & 10.78   & 10.92   & 10.77       & 11.03           & -0.008  & -0.077        & -4.856        \\
70.  & Hep-Th citations      & 27400   & 25.69   & 2468        & 323     & 106.82  & 95.75   & 88.46       & 105.40          & -0.030  & 0.591         & 1.012         \\
71.  & Cond-Mat 1995-2003    & 27519   & 8.44    & 202         & 125     & 38.30   & 28.36   & 26.52       & 21.29           & 0.166   & 0.234         & 0.339         \\
72.  & Digg                  & 29652   & 5.72    & 283         & 144     & 27.63   & 27.43   & 27.22       & 27.07           & 0.003   & 0.506         & 1.459         \\
73.  & Linux soft            & 30817   & 13.84   & 9338        & 302     & 154.98  & 137.60  & 69.53       & 851.62          & -0.175  & 3.741         & 3.186         \\
74.  & Enron                 & 33696   & 10.73   & 1383        & 334     & 115.48  & 108.16  & 90.59       & 141.36          & -0.117  & 1.720         & 2.447         \\
75.  & Hep-Ph citations      & 34401   & 24.46   & 846         & 302     & 74.33   & 63.34   & 61.91       & 62.50           & -0.006  & 0.037         & 0.245         \\
76.  & Cond-Mat 1995-2005    & 36458   & 9.42    & 278         & 159     & 49.17   & 37.21   & 34.58       & 26.88           & 0.177   & 0.264         & 0.398         \\
77.  & Gnutella Aug 30 2002  & 36646   & 4.82    & 55          & 50      & 11.39   & 10.09   & 9.93        & 10.46           & -0.104  & 0.075         & 0.225         \\
78.  & Slashdot              & 51083   & 4.56    & 2915        & 205     & 44.95   & 44.55   & 35.63       & 80.57           & -0.035  & 2.711         & 6.313         \\
79.  & Gnutella Aug 31 2002  & 62561   & 4.73    & 95          & 56      & 11.48   & 10.17   & 10.05       & 10.60           & -0.093  & -0.005        & 0.173         \\
80.  & Facebook              & 63392   & 25.77   & 1098        & 377     & 130.82  & 111.30  & 105.41      & 87.05           & 0.177   & 0.339         & 0.527         \\
81.  & Epinions              & 75877   & 10.69   & 3044        & 491     & 181.65  & 173.70  & 161.62      & 182.88          & -0.041  & 1.404         & 1.849         \\
82.  & Slashdot zoo          & 79116   & 11.82   & 2534        & 449     & 127.57  & 117.90  & 106.05      & 145.30          & -0.075  & 0.634         & 1.600         \\
83.  & Flickr                & 105722  & 43.83   & 5425        & 1003    & 614.42  & 504.27  & 572.00      & 348.21          & 0.247   & 0.763         & -1.909        \\
84.  & Wikipedia edits       & 113123  & 35.82   & 20153       & 1032    & 389.69  & 377.78  & 289.34      & 688.54          & -0.065  & 3.183         & 4.263         \\
85.  & Petster cats          & 148826  & 73.21   & 80634       & 2021    & 1160.43 & 1104.42 & 873.92      & 2195.71         & -0.164  & 4.215         & 3.265         \\
86.  & Gowalla               & 196591  & 9.67    & 14730       & 475     & 159.86  & 138.11  & 81.48       & 305.58          & -0.029  & 1.919         & 2.564         \\
87.  & Libimseti             & 220970  & 155.98  & 33389       & 3148    & 943.38  & 885.57  & 671.28      & 1639.96         & -0.139  & 2.355         & 3.098         \\
88.  & EU email              & 224832  & 3.02    & 7636        & 545     & 97.09   & 84.29   & 72.95       & 566.65          & -0.189  & 3.166         & 1.269         \\
89.  & Web Stanford          & 255265  & 15.21   & 38625       & 670     & 423.82  & 338.59  & 336.93      & 2029.74         & -0.116  & 2.262         & 0.038         \\
90.  & Amazon Mar 2 2003     & 262111  & 6.87    & 420         & 153     & 17.80   & 10.25   & 10.04       & 10.14           & -0.002  & 0.019         & 0.054         \\
91.  & DBLP collaborations   & 317080  & 6.62    & 343         & 199     & 114.72  & 70.14   & 112.00      & 20.75           & 0.267   & 0.796         & -5.594        \\
92.  & Web Notre Dame        & 325729  & 6.69    & 10721       & 438     & 175.66  & 173.02  & 164.69      & 279.68          & -0.053  & 4.893         & 2.851         \\
93.  & Math Sci Net          & 332689  & 4.93    & 496         & 162     & 33.53   & 20.74   & 23.00       & 15.43           & 0.103   & 0.098         & -0.390        \\
94.  & CiteSeer              & 365154  & 9.43    & 1739        & 441     & 52.54   & 33.69   & 35.33       & 47.45           & -0.063  & 0.045         & -0.182        \\
95.  & Zhishi                & 372840  & 12.43   & 127066      & 454     & 933.94  & 909.82  & 942.62      & -1737.11        & -0.282  & 6.198         & -2.044        \\
96.  & Actor coll net        & 374511  & 80.18   & 3956        & 2051    & 847.55  & 642.68  & 592.15      & 417.32          & 0.226   & 0.505         & 0.441         \\
97.  & Amazon Mar 12 2003    & 400727  & 11.73   & 2747        & 346     & 35.03   & 22.76   & 31.68       & 29.33           & -0.020  & 0.085         & -2.597        \\
98.  & Amazon Jun 6 2003     & 403364  & 12.11   & 2752        & 346     & 40.31   & 24.22   & 33.06       & 29.55           & -0.018  & 0.072         & -1.594        \\
99.  & Amazon May 5 2003     & 410236  & 11.89   & 2760        & 344     & 40.36   & 24.35   & 32.59       & 29.93           & -0.017  & 0.091         & -1.446        \\
100. & Petster dogs          & 426485  & 40.06   & 46503       & 2215    & 734.01  & 660.95  & 427.52      & 1803.39         & -0.088  & 2.297         & 2.868         \\
101. & Road network PA       & 1087562 & 2.83    & 9           & 9       & 3.11    & 2.25    & 2.90        & 2.20            & 0.122   & 0.000         & -2.818        \\
102. & YouTube friend net    & 1134890 & 5.27    & 28754       & 978     & 185.14  & 151.93  & 105.78      & 493.53          & -0.037  & 1.580         & 1.742         \\
103. & Road network TX       & 1351137 & 2.78    & 12          & 9       & 3.56    & 2.20    & 3.51        & 2.15            & 0.127   & 0.000         & -6.606        \\
104. & AS Skitter            & 1694616 & 13.09   & 35455       & 1739    & 653.66  & 361.86  & 260.77      & 1444.15         & -0.081  & 1.036         & 0.595         \\
105. & Road network CA       & 1957027 & 2.82    & 12          & 11      & 3.32    & 2.22    & 3.17        & 2.17            & 0.121   & 0.000         & -3.966        \\
106. & Wikipedia pages       & 2070367 & 40.90   & 230040      & 3611    & 775.44  & 640.35  & 308.90      & 1570.37         & -0.042  & 1.937         & 2.479         \\
107. & US Patents            & 3764117 & 8.77    & 793         & 370     & 110.45  & 31.76   & 75.82       & 20.34           & 0.168   & 0.001         & -1.641        \\
108. & DBpedia               & 3915921 & 6.42    & 469692      & 1738    & 462.92  & 389.51  & 388.33      & 1050.61         & -0.043  & 0.275         & 0.032         \\
109. & LiveJournal           & 5189808 & 18.76   & 15016       & 1636    & 537.93  & 268.64  & 408.16      & 154.42          & 0.039   & 0.033         & -1.460

\end{longtable*}